\documentclass[amsmath,amssymb,aps,prl,twocolumn]{revtex4-1}	

\usepackage{txfonts}
\usepackage{graphicx}
\usepackage{bm}
\usepackage{array}
\usepackage{color}
\usepackage[normalem]{ulem}

\newcommand{\ET}{$\kappa$-(BEDT-TTF)$_2$Cu$_2$(CN)$_3$}
\newcommand{\dmit}{EtMe$_3$Sb[Pd(dmit)$_2$]$_2$}

\newcommand{\Hcat}{$\kappa$-H$_3$(Cat-EDT-TTF)$_2$}

\begin{document}

\title{Presence and absence of itinerant gapless excitations in the quantum spin liquid candidate EtMe$_3$Sb[Pd(dmit)$_2$]$_2$}

\author{M. Yamashita$^1$}
 \email[]{my@issp.u-tokyo.ac.jp}
\author{Y. Sato$^2$, T. Tominaga$^2$, Y. Kasahara$^2$, S. Kasahara$^2$, H. Cui$^3$, R. Kato$^3$, T. Shibauchi$^4$, and Y. Matsuda$^2$}

\affiliation{$^1$The Institute for Solid State Physics, The University of Tokyo, Chiba 277-8581, Japan}
\affiliation{$^2$Department of Physics, Kyoto University, Kyoto 606-8502, Japan}
\affiliation{$^3$RIKEN, Saitama 351-0198, Japan}
\affiliation{$^4$Department of Advanced Materials Science, University of Tokyo, Chiba 277-8561, Japan}

\begin{abstract}
EtMe$_3$Sb[Pd(dmit)$_2$]$_2$, an  organic Mott insulator with nearly isotropic triangular lattice, is a candidate material for a quantum spin liquid, in which the zero-point fluctuations do not allow the spins to order.  The itinerant gapless excitations inferred from the thermal transport measurements in this system have been a hotly debated issue recently. While the presence of a finite linear residual thermal conductivity, $\kappa_0/T \equiv \kappa/T (T \rightarrow 0)$, has been shown [M. Yamashita {\it et al.} Science {\bf 328}, 1246 (2010)], recent experiments  [P. Bourgeois-Hope {\it et al.}, Phys. Rev. X {\bf 9}, 041051 (2019); J. M. Ni {\it et al.}, Phys. Rev. Lett. {\bf 123}, 247204 (2019)] have reported the absence of $\kappa_0/T$.  Here we show that the low-temperature thermal conductivity strongly depends on the cooling process of the sample.  When cooling down very slowly, a sizable $\kappa_0/T$ is observed.  In contrast, when cooling down rapidly, $\kappa_0/T$ vanishes and, in addition, the phonon thermal conductivity is strongly suppressed.  These results suggest that possible random scatterers introduced during the cooling process are responsible for the apparent discrepancy of the thermal conductivity data in this organic system.  The present results provide evidence that the true ground state of EtMe$_3$Sb[Pd(dmit)$_2$]$_2$ is likely to be a quantum spin liquid with itinerant gapless excitations.  
\end{abstract}

\maketitle

A quantum spin liquid (QSL) is a novel state of matter in which quantum fluctuations prevent the system from the long-range magnetic ordering even at zero temperature.  The ground states of the QSLs are quantum mechanically entangled and are expected to host fractional quasiparticle excitations~\cite{SavaryBalents2016,Zhou2017}.    In two-dimensional (2D) and 3D systems, it is widely believed that QSL ground states may emerge when interactions among the magnetic degrees of freedom are incompatible with the underlying crystal geometry.  
Typical 2D examples of such systems can be found in geometrically frustrated triangular and kagome lattices~\cite{SavaryBalents2016,Zhou2017} and in Kitaev honeycomb lattice with strong exchange frustration induced by the bond-directional interactions~\cite{Motome2020}.   
Among them, the spin-1/2 2D triangular-lattice Heisenberg antiferromagnet has been most widely discussed, because it has been suggested to be a prototype of a QSL in resonating valence bond (RVB) model with a quantum superposition of spin singlets~\cite{ANDERSON1973, FazekasAnderson1974}.  An important property of RVB liquids is that they have exotic excitations, called spinons, which are fractionalized quasiparticles that carry spin 1/2 but no charge.

The three organic Mott insulators {\ET}~\cite{Shimizu2003,Shimizu2006,SYamashita2008,MYamashita2008}, {\dmit}~\cite{Itou2008, Itou2010,MYama2010,SYamashita2011,Watanabe2012} and {\Hcat}~\cite{Isono2014,Shimozawa2017,SYamashita2017}
have been paid much attention recently as promising candidates to host a QSL state in real bulk materials with 2D triangular lattice.  In these organic systems, no magnetic ordering occurs at least down to low temperatures corresponding to 1/1000 of $J/k_{B}=200$--300\,K ($J$ is the exchange interaction between neighboring spins)~\cite{Shimizu2003, Itou2008,Isono2014}.  One of the most remarkable features commonly observed in these compounds is that the heat capacity has a non-zero linear-in-temperature term $\gamma T$, similar to metals, indicating the presence of gapless excitations despite gapped charge degrees of freedom~\cite{SYamashita2008,SYamashita2011,SYamashita2017}.  However, the nature of the gapless excitations has been highly controversial. The thermal conductivity  provides a powerful tool to probe elementary \emph{itinerant} excitations, which is totally insensitive to localized excitations such as those responsible for Schottky contributions that contaminate the heat capacity data at low temperatures.  The thermal conductivity is formed primarily by both acoustic phonons and itinerant spin  excitations, $\kappa=\kappa_{ph}+\kappa_{spin}$.  As $\kappa_{ph}/T$ vanishes at zero temperature, the observation of finite $\kappa_0/T\equiv \kappa/T (T\rightarrow 0)$ provides direct evidence for the presence of the itinerant gapless spin excitations.   Despite tremendous efforts, however, the true nature of the ground states of these organic QSL candidates has still remained elusive.

In {\Hcat}, a finite $\kappa_0 / T$ has been observed together with the concomitant emergence of QSL and quantum paraelectric properties at low temperatures~\cite{Shimozawa2017}.
It has been pointed out that this concomitant emergence may be realized by the coupling between the spin-1/2 residing in the (Cat-EDT-TTF)$_2$ dimers and the quantum fluctuations of protons bridging the layers of the 2D triangular lattice of the dimers. Therefore, the ground state may be different from QSLs expected in the simple 2D triangular lattice.
In {\ET} and {\dmit}, the 2D layers of the triangular (BEDT-TTF)$_2$ or [Pd(dmit)$_2$]$_2$ dimers are well separated by nonmagnetic layers, forming a slightly distorted 2D triangular lattice.
In {\ET}, a vanishingly small  $\kappa_0/T$ has been reported~\cite{MYamashita2008}.
However, it has been suggested that {\ET} has an inhomogeneous and phase separated spin state~\cite{Shimizu2006,Nakajima2012}.
Therefore, it is still ambiguous whether the absence of  $\kappa_0/T$  is an intrinsic property of the uniform QSL state of the organic Mott insulator.

It has been pointed out that {\dmit} may be more suitable to single out genuine features of the QSL, compared with {\ET}, because a more homogeneous QSL state is attained at low temperatures~\cite{Itou2010}. 
However, the low-temperature thermal conductivity of {\dmit} has  been highly controversial.  A sizable $\kappa_0/T$ term has been reported in Ref.~[\onlinecite{MYama2010}], which indicates the presence of highly itinerant and gapless excitations.  These results, together with the results of Pauli-paramagnetic-like low energy excitations revealed by magnetic torque measurements, have been discussed in terms of the spinons that form the Fermi surface~\cite{Watanabe2012}.  On the other hand, recent thermal conductivity measurements by two groups have reported the vanishingly small $\kappa_0/T$~\cite{BourgeoisHope2019,NiSYLee2019}.   Moreover, it has been reported that $\kappa_{ph}$ is also strikingly suppressed from that reported in Ref.~[\onlinecite{MYama2010}], suggesting that the phonon mean free path $\ell_{ph}$ becomes extremely short at low temperatures, comparable to that of amorphous materials~\cite{Yamashita_2019_DMIT}.  The absence of itinerant spin excitations and strong suppression of $\ell_{ph}$  have been discussed in terms of the strong coupling between spin excitations and phonons~\cite{BourgeoisHope2019,NiSYLee2019}.

The presence or absence of itinerant spin excitations is intimately related to the intrinsic properties of the QSL state, such as possible spinon Fermi surface. It is therefore of primary importance to clarify the origin of the apparent discrepancy of the low-temperature thermal conductivity results.  Here, we report on measurements of the low-temperature thermal conductivity after cooling the crystals from the room temperature at different cooling rates, ranging from $-0.4$\,K/h to $-13$\,K/h.   
Our major finding is that a finite $\kappa_0/T$ is clearly resolved in the slow-cooling condition, whereas it is absent in the rapid-cooling condition.  
Moreover, in rapidly cooled crystals, we find that phonon thermal conductivity is strongly suppressed, indicating a large enhancement of the phonon scattering. 
These results reveal that the low-temperature transport properties of {\dmit} do depend on the cooling process of the crystal, suggesting that possible random scatterers introduced during the cooling process are responsible for the apparent discrepancy of the thermal conductivity data.  The present results indicate that the QSL ground state of {\dmit} is likely to have intrinsically itinerant gapless excitations.

We have measured the temperature dependence of thermal conductivity of three single crystals by using the standard steady-state method.
These single crystals were taken from the same batch used in the previous report (E and F in Ref.~\cite{Yamashita_2019_DMIT}).
The thermal contacts to the thermometers and the heater were made by attaching gold wires to the sample by carbon paste. 
These setups were essentially the same with those used in the previous report~\cite{MYama2010,Yamashita_2019_DMIT}.
To ensure the validity of the experimental setup, the reproducibility of results, and the accuracy of the absolute value of $\kappa$ at low temperatures, we have also measured $\kappa$ of a stainless steel with a similar thermal resistance as the present {\dmit} crystals in the same setup and confirmed the verification of the Wiedemann-Franz law. 
These {\dmit} crystals were cooled down with different cooling rates of approximately $-0.4$\,K/h (\#1), $-1.5$\,K/h (\#2), and $-13$\,K/h (\#3) from the room temperature to $\sim4$\,K.  Then they were cooled down to the lowest temperatures.
Here we denote the cooling rates of $-0.4$ and $-1.5$\,K/h as slow cooling and $-13$\,K/h as rapid cooling.
After the measurement, each crystal was analyzed by X-ray diffraction (XRD) method, which indicated high crystal quality with $R$-factors of 2--3\%. The XRD measurements at the room temperature found no difference between these three samples. This was also the case of the previous report~\cite{BourgeoisHope2019}. 
Because the crystals often do not recover to the initial state after a thermal cycle, we discuss the data taken at the first cooling from the room temperature.

\begin{figure}[t]
\centering
\includegraphics[width=\linewidth]{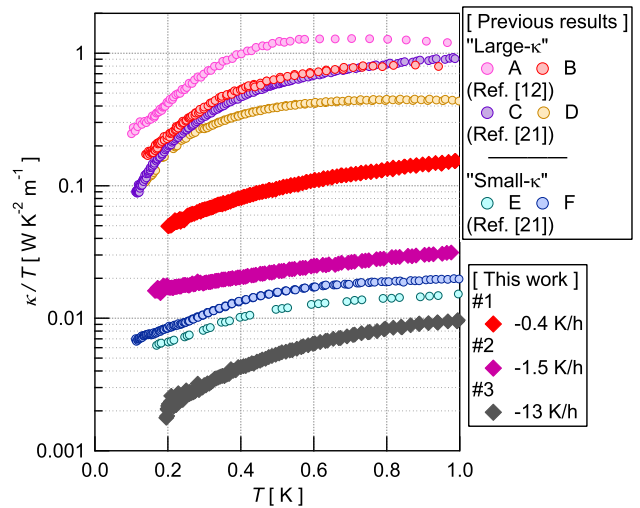}
\caption{
	(Color online) The temperature dependence of the thermal conductivity divided by the temperature of {\dmit}. Crystal \#1, \#2 and \#3 were cooled down by approximately $-0.4$ (red diamonds), $-1.5$ (purple diamonds), and $-13$\,K/h (gray diamonds), respectively. 
	The data of A to F (from Ref.~\cite{MYama2010,Yamashita_2019_DMIT}) is also plotted for a comparison.
	The cooling rate of A to F was approximately -10\,K/h (A to D), -100\,K/h (E), and -30\,K/h (F).
}
\label{k_T_log}
\end{figure}

Figure~\ref{k_T_log} depicts the temperature dependence of the thermal conductivity in zero field for Crystal \#1, \#2 and \#3, along with the previously reported data (A--F from Ref.~\cite{MYama2010,Yamashita_2019_DMIT}).  
Since the thermal conductivity is dominated by $\kappa_{ph}$ at relatively high temperatures around $\sim$1\,K, the previous results can be divided into two groups; the data for A, B, C, and D exhibit large $\kappa_{ph}$ value at high temperatures and finite $\kappa_0/T$ (''large-$\kappa$ group"), while E and F show small $\kappa_{ph}$ and vanishingly small $\kappa_0/T$ (''small-$\kappa$ group").  Obviously, $\kappa/T$ data for Crystals \#1 and \#2 measured after the slow cooling (red and purple diamonds) have intermediate values between the two groups.  
For Crystal \#3, $\kappa/T$ data is even below that of the small-$\kappa$ group.

\begin{figure}[!tbh]
	\centering
	\includegraphics[width=\linewidth]{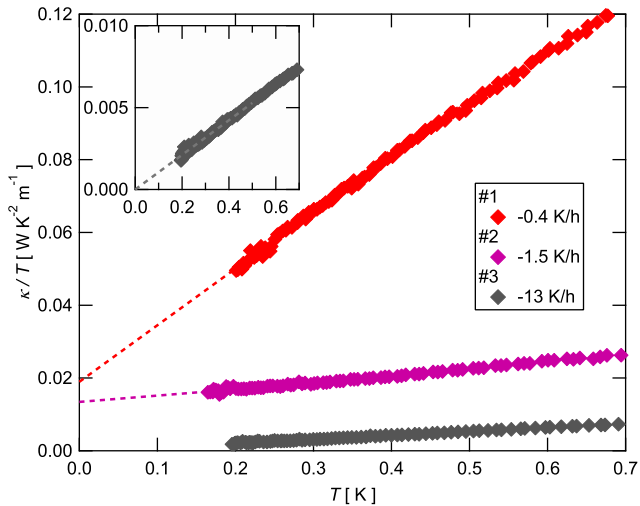}
	\caption{(Color online)
		The low-temperature data of $\kappa/T$ plotted as a function of $T$. The dashed lines show an extrapolation of the data. The inset shows an enlarged view of the data of Crystal \#3.
	}
	\label{k_T_LT}
\end{figure}

The cooling rate is found to have two effects on the low-temperature thermal conductivity.   As shown in Fig.~\ref{k_T_LT} and its inset, $\kappa/T$ exhibits nearly $T$-linear dependence at very low temperatures.  In the slow cooling data shown by red and purple diamonds in Fig.~\ref{k_T_LT}, the linear extrapolation of $\kappa/T$ to $T\rightarrow 0$ exhibits finite intercepts for both Crystal \#1 and \#2, indicating the presence of the residual $\kappa_0/T$ term.  We note that the magnitude of $\kappa_0/T$ for both crystals is close to that of C and D in the previous report~\cite{Yamashita_2019_DMIT}. 
In the rapid cooling data, in stark contrast, the linear extrapolation of $\kappa/T$ to $T\rightarrow 0$ for Crystal \#3 indicates vanishingly small $\kappa_0/T$ as shown in the inset of Fig.~\ref{k_T_LT}. 
These results indicate that the gapless itinerant excitations appear in the slow cooling, while they disappear in the rapid cooling condition.

In the crystals where $\kappa_0/T$ is vanishingly small, the thermal conductivity at finite temperatures, which is dominated by the phonon contribution, is also largely suppressed. 
The magnitude of $\kappa$ at finite temperatures is determined by the phonon mean path.
For example, $\kappa$ of a crystalline SiO$_2$ is 2--3 orders of magnitude larger than that of a vitreous SiO$_2$ at $\sim 1$\,K~\cite{ZellerPohl1971}. In {\dmit}, $\ell_{ph}$ of the small-$\kappa$ group is comparable to that of amorphous materials~\cite{Yamashita_2019_DMIT}.  
This extremely short $\ell_{ph}$ has been discussed in terms of intrinsically strong spin-phonon scattering in Ref.~\cite{BourgeoisHope2019}.  However, the present results, which show the strong cooling-rate dependence of $\kappa_{ph}$, clearly imply that $\ell_{ph}$ is strikingly suppressed by random scatterers induced by the rapid cooling, excluding the possibility of the strong spin-phonon scattering.   

Recently, the low-temperature thermal conductivity measurements have been performed on transition metal dichalcogenide 1T-TaS$_2$~\cite{Murayama2020}, which is a candidate material that hosts a QSL ground state with spin-1/2 on the 2D perfect triangular lattice. In very clean samples of 1T-TaS$_2$,  the linear temperature term of the heat capacity and the finite  $\kappa_{0}/T$ are clearly resolved, consistent with the presence of gapless spinons with a Fermi surface. Introduction of additional weak random exchange disorder in 1T-Ta(S$_{1-x}$Se$_x$)$_2$ leads to vanishing of $\kappa_0/T$.  Moreover, YbMgGaO$_4$, which has a structurally perfect 2D triangular lattice and effective spin-1/2 local moments of Yb$^{3+}$ ions,  has also been suggested to be a QSL candidate.   The absence of $\kappa_0/T$ has been reported in YbMgGaO$_4$~\cite{YXu2016}.  In this compound, Mg$^{2+}$/Ga$^{3+}$ site mixing introduces strong randomness in the exchange interaction between spin-1/2 on Yb sites. These results provide strong support that the itinerant gapless excitations are sensitive to disorder.

It has been well known that physical properties of organic compounds are often sensitively altered by their cooling conditions.
For example, different ground states, from a superconducting or a metallic state to an insulating state, have been observed by changing the cooling rate in (TMTSF)$_2$$X$O$_4$ ($X$ = Cl~\cite{Takahashi1982}, Br~\cite{Tomic1984}), $\kappa$-(BEDT-TTF)$_2$Cu[N(CN)$_2$]Br~\cite{Su1998, Taniguchi1999}, and $\theta$-(BEDT-TTF)$_2X$(SCN)$_4$ ($X$ = RbZn~\cite{Nad2007}, TlCo, and CsZn~\cite{Sato2014}).
These different ground states are discussed in terms of the effects of disorder introduced by rapid cooling~\cite{Su1998, Taniguchi1999}, which may give rise to a significant suppression of thermal conductivity.
In fact, decrease of the thermal conductivity by disorderly domains of anion ordering in rapid cooled samples have been observed in (TMTSF)$_2$ClO$_4$~\cite{Belin1997}.
In {\dmit}, a large change of 15\% in the transfer integrals from the room temperature to 5\,K has been reported as a result of large contractions of the lattice constants~\cite{Ueda2018}.
Also, the vibrational spectroscopy measurements~\cite{Yamamoto2017} have reported that there are multiple modes of charge and lattice fluctuations.
These results imply that some inhomogeneous electronic states might be formed in a rapid cooling process, which can act as the random scatterers for the thermal conduction.

As suggested in Ref.~\cite{Yamashita_2019_DMIT},  in all A-D samples,   $\ell_{ph}$ becomes comparable to the effective sample size at very low temperatures.   On the other hand, $\ell_{ph}$ in Crystal \#1 and \#2 cooled down in the slow rate, which show the finite $\kappa_0/T$ comparable to those of C and D, is still below the effective sample size.  This may be due to the macroscopic domain structure of Crystals \#1 and \#2, which reduces the effective sample size. In this case, intrinsic $\kappa_0/T$ may be larger than the observed one.  The cooling rate of all samples A-D with a finite $\kappa_0/T$ is $\sim -10$\,K/h.  This cooling rate is comparable to that of Crystal \#3, in which $\kappa_0/T$ is absent. This conflicting result, i.e. the presence and absence of $\kappa_0/T$ in the similar cooling rate,  suggests that there may be another factor other than the cooling rate, which determines the formation of the random scatterers that suppresses both $\kappa_{ph}$ and $\kappa_0/T$. 
Clarifying all factors relating to the formation of the random scatterers remains as a future issue.

In summary, we have measured the low-temperature thermal conductivity of {\dmit} cooled down in a wide range of cooling rates. The low-temperature thermal conductivity strongly depends on the cooling rate.  A  finite linear-in-temperature thermal conductivity is clearly resolved under the slow-cooling conditions, whereas it is absent under the rapid-cooling conditions.  Moreover, a striking enhancement of the phonon scattering occurs  in the rapid cooled crystals. The present results suggest that the apparent discrepancy of the low-temperature thermal conductivity data in this organic system between different groups is attributed not to the sample dependent crystal quality but to random scatterers introduced during the cooling process.  These results provide support on the presence of highly mobile gapless excitations in a QSL candidate {\dmit}. 

\begin{acknowledgments}
We thank T. Sasaki for fruitful discussions.
This work is supported by Grants-in-Aid for Scientific Research (KAKENHI) (Nos. 15KK0160, 16H06346, 18H01177, 18H01180, 18H05227, 19H01848, 19K21842) and on innovative areas ”Topological Material Science” (No. 15H05852) ”Quantum Liquid Crystals” (No. 19H05824) from the Japan Society for the Promotion of Science, and JST CREST (JPMJCR19T5). 
\end{acknowledgments}

\providecommand*\hyphen{-}

\end{document}